\documentstyle[aps,prl,psfig,floats]{revtex}
\begin{document}
\draft
\twocolumn[\hsize\textwidth\columnwidth\hsize\csname@twocolumnfalse\endcsname

\title{Phonon Softening and Elastic Instabilities\\
in the Cubic-to-Orthorhombic Structural Transition of CsH}

\author{A.M. Saitta,$^1$\cite{e-mail} D. Alf\`e,$^1$\cite{e-mail}
S. de Gironcoli,$^1$\cite{e-mail} and S. Baroni$^{1,2}$\cite{e-mail} }

\address{$^1$INFM -- Istituto Nazionale per la Fisica della Materia
and \\ SISSA -- Scuola Internazionale Superiore di Studi Avanzati,
Via Beirut 2-4, I-34014 Trieste, Italy \\ $^2$CECAM -- Centre
Europ\'een de Calcul Atomique et Mol\'eculaire, ENSL, 46~All\'ee
d'Italie, 69364 Lyon Cedex 07, France} 

\date{\today}
\maketitle

\begin{abstract} The cubic-to-orthorhombic structural transition
occurring in CsH at a pressure of about 17 GPa is studied by {\it ab
initio} calculations. The relative stability of the competing
structures and the transition pressure are correctly predicted. We
show that this pressure-induced first-order transition is intimately
related to a displacive second-order transition which would occur upon
application of a shear strain to the (110) planes. The resulting
instability is rationalized in terms of the pressure-induced
modifications of the electronic structure. \end{abstract}

]
\narrowtext

In recent years the availability of the diamond anvil-cell technology
\cite{rassegna} has renewed the interest in the high-pressure
properties of alkali halides which are the simplest and prototypical
among ionic solids. In the case of cesium halides, these studies have
been carried out especially in the search of band-overlap
metallization which occurs at pressures in the Mbar range. In the
quest of metallization, unexpected phase transformations to tetragonal
and orthorhombic structures have been observed experimentally
\cite{mbn1,mbn2,mbn3,mbn4,mbn5} and studied theoretically
\cite{mbn5,sambapati`,mbn6,mbn7}.

Metal hydrides are considerably more covalent than the corresponding
halides. In fact, transition-metal hydrides are considered to be
covalent materials, whereas alkali hydrides are structurally rather
similar to the corresponding halides. Understanding the behavior of
these materials at high pressure would presumably allow to better
characterize the bonding properties of hydrogen in extreme pressure
conditions and thus provide valuable hints towards the long searched
goal of hydrogen metallization.

Alkali hydrides at zero pressure crystallize in the rocksalt (cubic
$B1$) structure, while most of them undergo a transition to the cubic
$B2$ (CsCl-like) structure at an applied pressure of a few GPa
\cite{ruo1,ruo2,ruo3}. Recently, a second transition from the $B2$
structure to a new orthorhombic phase has been observed to occur in
CsH at an applied pressure of about 17 GPa \cite{ruo4}. The new phase
has been assigned the CrB structure which belongs to the $D^{17}_{2h}$
space group.

In order to assess the driving mechanisms of the transition we have
undertaken a series of first-principles calculations of the
electronic, structural, elastic, and vibrational properties of CsH as
well as of their dependence upon pressure. To this end, we have
employed density-functional-theory (DFT) within the local-density
approximation (LDA) \cite{lda}, and its gradient-corrected (GC)
generalizations \cite{ggc}. Our calculations have been performed using
norm-conserving pseudopotentials \cite{pp}. In the case of Cs, $5s$
and $5p$ semi-core states have been treated as valence states. For
LDA calculations we have used the same Cs potential as in
Ref. \cite{mbn6}, while Hydrogen is described by a norm-conserving
pseudopotential constructed for the $1s$ wave-function, so as to
smoothen it somewhat and make our calculations well converged with the
25 Ry kinetic-energy cutoff which we use. New potentials have been
generated for GC calculations. The vibrational properties are
determined using the density-functional perturbation theory (DFPT)
described in Ref. \cite{bgt}.

\begin{table}
\caption{Comparison between calculated and experimentally observed
structural properties of CsH. $a_0$ (a.u.) is the equilibrium lattice
parameter in the $B1$ phase, $B_0$ (GPa) the bulk modulus, $P^*_1$ and
$P^*_2$ the transition pressures (GPa) to the $B1$ and CrB structures 
respectively, while $\Delta V_1$ and $\Delta V_2$ are the corresponding volume
changes. $V^*_2/V_0$ is the ratio of the volume at the $ B2
\rightarrow {\rm CrB}$ transition to the equilibrium volume.}
\begin{tabular}{lccccccc} & $a_0$ & $B_0$& $P^*_1$ & $\Delta V_1$ &
$P^*_2$ & $\Delta V_2$ & $V^*_2/V_0$ \\ \hline LDA & 11.62 & 11.3 &
-0.8 & 18.0\% & 11 & 4.5\% & 0.63 \\ GC & 12.25 & 10.3 &
\phantom{-}1.5 & 12.5\% & 15& 4.0\% & 0.54\\ Expt.\tablenotemark[1] &
12.07 & 8.0& \phantom{-}0.8 & \phantom{1}8.4\% & 17 & 6.3\% & 0.53\\
\end{tabular} \tablenotetext[1]{Ref. \cite{ruo4}} \label{t:1} 
\end{table}

In Table~\ref{t:1} we report our results for some structural
properties of CsH in the three phases studied here: $B1$, $B2$, and
CrB. The LDA fails to account for the stability of the rocksalt
structure at zero pressure (negative $B1\rightarrow B2$ transition
pressure), while GC calculations predict the correct sequence of
transitions. GC-DFT also considerably improves the agreement between
calculated and observed equilibrium lattice parameters and bulk
moduli. The inclusion of Cs semi-core states in the valence manifold
implies that the pseudopotential transferability is optimal around the
corresponding orbital energies, rather than at the energies of the
valence states. As a consequence, this treatment of semi-core
states---although essential to obtain sensible results
\cite{mbn6}---might require the use of more than one reference state
in order to ensure optimum transferability. Whether or not the
relatively poor quality of the LDA predictions is due to deficiencies
in the Cs pseudopotential and the improvements achieved by using
GC-DFT due to a fortuitous cancellation of errors is a matter which
deserves further investigations. In any events, our results seem to
confirm the importance of the gradient corrections to the LDA in the
description of structural phase transitions, recently claimed in the
case of the diamond to $\beta$-tin transition in silicon
\cite{Scheffler}. Both LDA and GC-DFT calculations correctly predict
that a transition from the $B2$ to the CrB phases would occur at
pressures somewhat larger than 10 GPa. In this case too, GC-DFT gives
a value of the transition pressure in closer agreement with
experiments (15 {\it vs.} 17 GPa). Note however that---due to
hysteresis effects---the experimental estimate done when loading the
sample \cite{ruo4} only provides an upper limit to the transition
pressure. Also the volume discontinuity and the ratio between the
transition volume and the equilibrium one are well reproduced by our
calculations.

\begin{figure} \centerline{\psfig{file=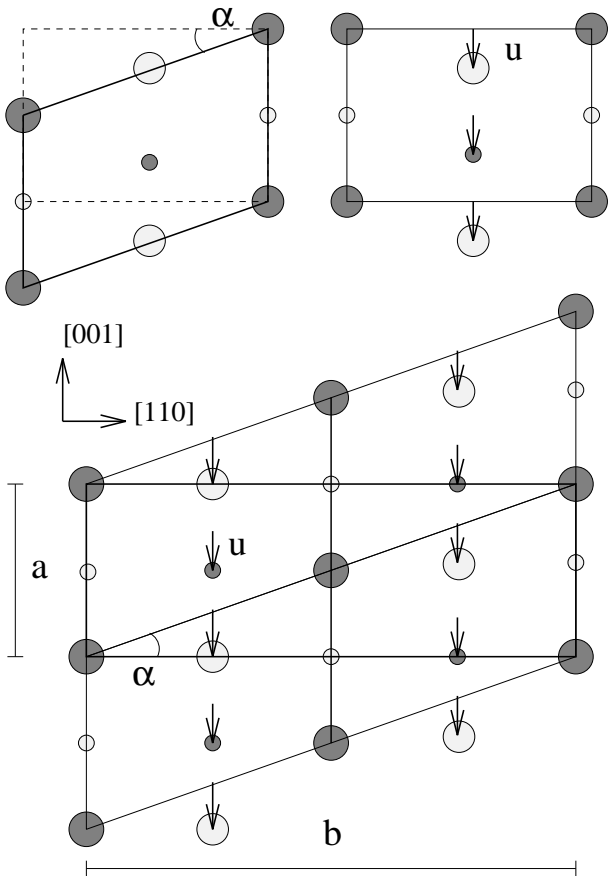,width=8cm}}
\caption{Decomposition of the distortion leading to the CrB structure
(lower panel) into an elastic (upper left) and a vibrational (upper
right) contributions. Large and small circles represent Cs and H atoms
respectively. Dark and light shading refer to the topmost and second
(110) layers respectively} \label{f:struct} \end{figure}

The CrB structure can be viewed as resulting from a continuous
deformation of the cubic $B2$ structure in which a shear of the cubic
edges in the (110) planes is coupled to a vibrational distortion of
the atomic lattice, having the periodicity of a zone-border transverse
phonon at the $M$ point of the Brillouin zone (BZ), ${\bf q} =
({1\over 2}, {1\over 2}, 0)$, with symmetry $M^-_2$, and polarized
along $(001)$. 
In Fig. 1 we display how the distorted (110) plane
(lower panel) results from the combination of the elastic (upper
left) and the vibrational (upper right) deformations. In the ideal CrB
structure the $a$ and $b$ crystallographic parameters indicated in
Fig. 1 are in the ratio $b/a=2\sqrt{2}$ which would be appropriate to
a cubic structure. Due to the lower symmetry of the CrB structure with
respect to the cubic one, the actual value of $b/a$ slightly differs
from the ideal one and depends somewhat on pressure. This dependence
is however very weak: at the transition pressure, for instance, one
has $b/a\approx 2.9$. The third crystallographic axis is orthogonal to
the $ab$ plane, and the corresponding crystallographic parameter has
an ideal value $c=\sqrt{2}a$. The pressure dependence of $c/a$ is
somewhat stronger than that of $b/a$: at the transition, for instance,
one has $c/a \approx 1.3$.

\begin{figure} \centerline{\psfig{file=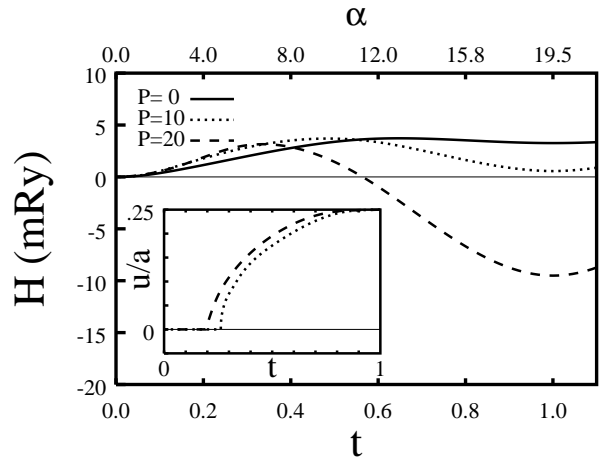,width=8cm}}
\caption{Enthalpy {\it vs.} angle of deformation for different values
of the applied external pressure. In the inset we show the
correponding values of the amplitude of the lattice distortion, $u$.}
\label{f:enth}
\end{figure}

In Fig. 2 we show the $T=0$ enthalpy, $H=E+PV$, of the system
constrained to a given elastic deformation, as a function of its
amplitude which we measure through the parameter $t=\tan(\alpha)
\times b/a$ \cite{lda-analysis}. We see that at any positive pressure
a second minimum exists for $t=1$, corresponding to the CrB
orthorhombic structure. Actually, the exact value of $\alpha$
slightly depends on pressure because so does $b/a$. For pressures
higher than $P_2^\star \approx 11~\rm GPa$ the second minimum becomes
more stable giving rise to a first-order phase transition. The
amplitude of the vibrational distortion, $u\equiv x/a$, which
minimizes the enthalpy, $u_{min}$, depends on $t$, but for $t=1$ it is
constrained by symmetry to be $u_{min}=1/4$, independent of
pressure. The dependence of $u_{min}$ upon the amplitude of the
elastic distortion, $t$, is shown in the inset of Fig. 2. We see that
$u_{min}$ remains equal to zero up to a critical value of $t=t^\star$
which depends on pressure, and that it saturates to $1/4$ for
$t=1$. This behavior is typical of the order parameter at a
second-order phase transition, and it indicates therefore that a
phonon frequency softens when the elastic distortion becomes larger
than $t^\star$. In Fig. 3 we display the $M^-_2$ phonon frequency as a
function of $t$ at different pressures, $P$, and as a
function of $P$ for different elastic distortions.

\begin{figure}
\twocolumn[\hsize\textwidth\columnwidth\hsize\csname@twocolumnfalse\endcsname
\centerline{\psfig{file=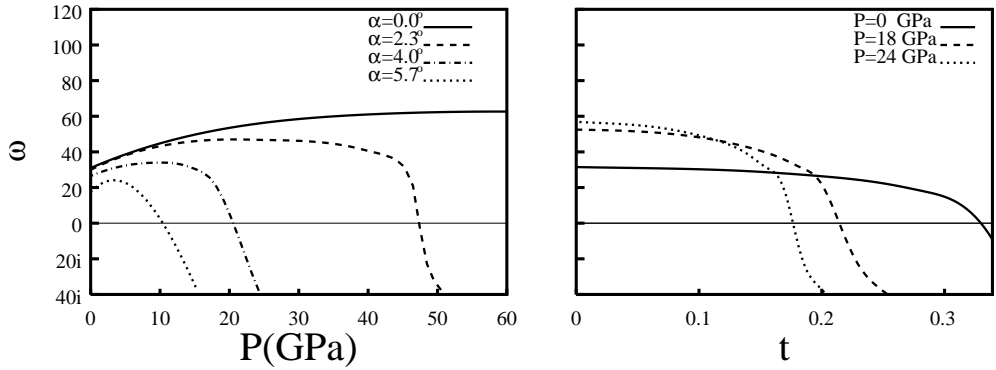,height=5cm}} \caption{Frequency of the
$M_2^-$ phonon mode (cm$^{-1}$) as a function of the applied
hydrostatic pressure for different angles of deformation of the cubic
structure (left), and as a function of the angle of deformation for
different pressures (right).} \label{f:soft} ] \end{figure}

The softening of the $M^-_2$ frequency in correspondence to some value
of the elastic distortion could have been expected on the basis of
symmetry considerations. To see this, let us consider the crystal
energy as a function of $t$ and $u$, $E(t,u)$. Crystal symmetry
requires that $E(t,u) = E(\pm t,u) = E(t,\pm u) = E(t+2,u+{1\over 2})
= E(t+2,u-{1\over 2})$. These relations imply that $E(t,u)$ is
stationary at the point $(t,u)=(2,0)$. This point lies in between the
four equivalent minima at $(0,0)$, $(2,\pm{1\over 2})$, and
$(4,0)$. Assuming that the energy landscape is simple, we arrive at
the conclusion that $(2,0)$ is a maximum. In particular, one has that
$\partial^2 E / \partial u^2 < 0 $, and hence there exists a value of
$0< t < 2$ for which $\omega_{M_2^-}^2 \sim \partial^2 E / \partial
u^2 =0$. The orthorhombic phase corresponds to the $(1,{1\over 4})$
point in the $(t,u)$ plane which is also stationary because of
symmetry. This point is located halfway between the two points $(1,0)$
and $(1,{1\over 2})$ which are equivalent by symmetry and stationary
with respect to variations of $u$ ($\partial E /\partial u=0$).

If $t^\star$ were larger than 1, $\partial^2E/\partial u^2$ would be
positive at $(1,0)$ and $(1,{1\over 2})$, and therefore the assumption
of a simple energy landscape would imply that $(1,{1\over 4})$ is a
maximum with respect to $u$, and the orthorhombic structure
unstable. It is easy to see that a first-order transition with a
discontinuous variation of $u$ would occur in this case as the elastic
distortion crosses $t=1$. In fact, in this case the line $(t,0)$ is
stable with respect to variations of $u$, up to $t=1$. For $t>1$ one
has that $E(t,{1\over 2}) < E(t,0)$. Hence, a discontinuous jump in $u$
would occur at $t=1$ where $E(t,{1\over 2}) = E(t,0)$.

Arguments similar to those expounded above show that for $t^\star < 1$
the $(1,{1\over 4})$ point is a minimum in the $u$ direction, while
nothing can be concluded in principle for other directions, but the
fact that local stability is allowed. Our calculations indicate that
in the present case $t^\star<1$ and that the CrB structure is indeed
locally stable. For pressures higher than 11 GPa (15 GPa within GC),
this structure is favored with respect to the cubic one, giving thus
rise to a first-order transition.

The mechanisms driving the softening of the $M_2^-$ phonon are related
to the metallization of CsH under an applied pressure and/or shear
deformation. At equilibrium ($P=0$, $t=0$), the maximum of the valence
band is practically degenerate between the $R\equiv({1\over 2}{1\over
2}{1\over 2})$ and $X\equiv (00{1\over 2})$ points of the BZ, and the
minimum of the conduction bands lies at the $R$ point. A small applied
pressure lifts the degeneracy between the
valence-band maxima, rising the $X$ point with respect to $R$. The
resulting fundamental band gap, $E_g$, corresponds therefore to a
$X\rightarrow R$ transition, whose transferred momentum is $M\equiv
({1\over 2}{1\over 2}0)$ , {\it i.e.} the wave-vector of the phonon
which goes soft. 

In Fig. 4 we display the electronic energy bands of
CsH at $P=17.5~{\rm GPa}$, for $t=0$ and for $t=0.3$. In the
distorted structure---corresponding to a non-vanishing value of $t$---the
$R$ and $X$ points are folded to a same point of the BZ, and so are
the $\Gamma$ and $M$ points. The gap, which is indirect in cubic
symmetry, becomes thus direct in the distorted structure. We see that
the elastic distortion considerably favors metallization, and thus
enhances the screening of the phonon frequencies at the $M$ point of
the cubic BZ. 

In Fig. 5 we display the dependence of $E_g$ upon
pressure and deformation angle. At small pressures, the $M_2^-$ phonon
frequencies softens almost for the same values of $t$ for which the
gap closes, whereas for pressure larger than $\approx 18~{\rm GPa}$,
it takes a larger value of $t$ to soften the phonon frequency than to
close the gap.
If the bands were parabolic around the valence- and
conduction-band edges, the phonon would go soft at the
metallization. In fact, the independent-electron polarizability,
$\chi_0({\bf q}_M)$ would diverge logarithmically in this case
\cite{halperin}, and the restoring force for a lattice distortion of
wave-vector ${\bf q}_M$ would correspondingly vanish. If the valence
and conduction bands have different shapes, the logarithmic divergence
would no longer hold, and all that can be said in this case is that
the more these shapes are similar, the more metallization contributes
to the phonon softening. In any events, when the amplitude of the
elastic distortion is larger than $t^\star$, the spontaneous
vibrational distortion due to phonon softening re-opens the band gap
and makes the orthorhombic phase insulating, in agreement with the
behavior of $E_g$ found in~\cite{ruo5}.

\begin{figure}
\twocolumn[\hsize\textwidth\columnwidth\hsize\csname@twocolumnfalse\endcsname
\centerline{\psfig{file=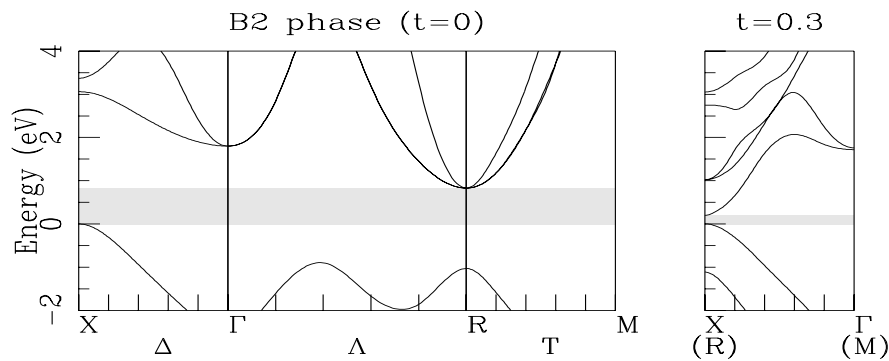,height=6cm,width=12cm}} \caption{Electron energy
bands of CsH close the fundamental energy gap (which is evidentiated
by the shadowed area) at an applied pressure of 17.5~GPa in the cubic
structure(left) and under a finite shear deformation (right)}
\vskip 8pt
\label{f:bands} ]\end{figure}

\begin{figure}
\twocolumn[\hsize\textwidth\columnwidth\hsize\csname@twocolumnfalse\endcsname
\centerline{\psfig{file=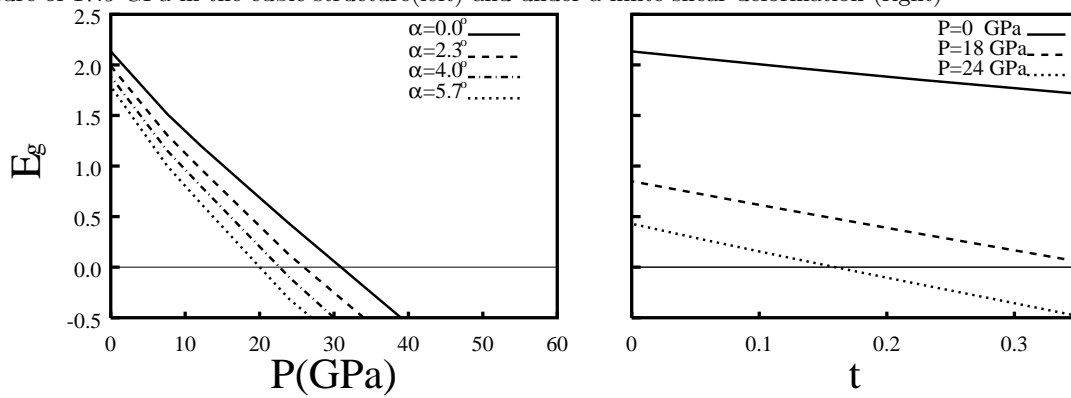,width=14cm}}\caption{Fundamental
energy gap (eV) as a function of the applied hydrostatic pressure for
different angles of deformation of the cubic structure (left), and as
a function of the angle of deformation for different pressures
(right). }
\vskip 8pt \label{f:gap} ] \end{figure}

We would like to thank J.-M. Besson and E. Tosatti for very useful
discussions.


\begin{references}
\bibitem{lda-analysis} 	
	The analysis which follows is based on calculations performed
	within the LDA. Our conclusions, of course, do not depend on
	the form of the exchange-correlation functionals.
\end{references}
\end{document}